\documentclass[preprint,10pt]{elsarticle}

\usepackage{amssymb}
\usepackage{graphicx}
\usepackage{wrapfig}
\usepackage{amsmath,amssymb}
\usepackage{amsthm}
\usepackage{bm}
\usepackage{setspace}
\usepackage{amsfonts}
\usepackage{ascmac}
\usepackage{varioref}
\usepackage{hhline}
\usepackage{color}
\usepackage{multirow}
\usepackage{algpseudocode, algorithm}
\usepackage{caption}

\usepackage{stmaryrd}
\begin{document}

\onehalfspacing 
\def\E{\mathop{{\rm E}}}
\def\Var{\mathop{{\rm Var}}}
\def\etr{\mathop{{\rm etr}}}
\def\tr{\mathop{{\rm tr}}}
\def\diag{\mathop{{\rm diag}}}
\def\b#1{\mathbf{#1}}
\def\bs#1{\mbox{$\boldsymbol#1$}}

\theoremstyle{plain}
\newtheorem{thm}{Theorem}
\newtheorem{cor}[thm]{Corollary}
\newtheorem{prop}{Proposition}
\newtheorem{lem}{Lemma}

\theoremstyle{definition}
\newtheorem{remark}{Remark}

\newcommand{\argmax}{\mathop{\rm arg~max}\limits}
\newcommand{\argmin}{\mathop{\rm arg~min}\limits}
\renewcommand{\multirowsetup}{\centering}
\newcommand{\indep}{\mathop{\perp \! \! \! \perp}}
\begin{frontmatter}



\title{
Test for mean matrix in GMANOVA model under heteroscedasticity and non-normality for high-dimensional data}


\author{Takayuki Yamada
\corref{cor1}}
\ead{takayuki-yamada@riko.shimane-u.ac.jp}
\address{
Department of Mathematical Sciences, \\
Shimane University, \\
1060 Nishikawatsu-cho, Matsue, Shimane 690-8504, Japan}
\cortext[cor1]{Corresponding author}
\author{Tetsuto Himeno
}
\address{
Faculty of Data Science, \\
Shiga University,\\
1-1-1 Banba, Hikone, Shiga 522-8522, Japan}
\author{Annika Tillander
}
\address{
Department of Computer and Information Science, \\
Link\"{o}ping University,\\
581 83 Link\"{o}ping, Sweden}
\author{Tatjana Pavlenko
}
\address{
Department of Mathematics, \\
KTH Royal Institute of Technology, \\
SE-100 44 Stockholm, Sweden}

\begin{abstract}
This paper is concerned with the testing bilateral linear hypothesis on the mean matrix in the context of the generalized multivariate analysis of variance (GMANOVA) model when the dimensions of the observed vector may exceed the sample size, the design may become unbalanced, the population may not be normal, or the true covariance matrices may be unequal. 
The suggested testing methodology can treat many problems such as the one- and two-way MANOVA tests, the test for parallelism in profile analysis, etc., as specific ones. 
We propose a bias-corrected estimator of the Frobenius norm for the mean matrix, which is a key component of the test statistic. 
The null and non-null distributions are derived under a general high-dimensional asymptotic framework that allows the dimensionality to arbitrarily exceed the sample size of a group, thereby establishing consistency for the testing criterion. 
The accuracy of the proposed test in a finite sample is investigated through simulations conducted for several high-dimensional scenarios and various underlying population distributions in combination with different within-group covariance structures. 
Finally, the proposed test is applied to a high-dimensional two-way MANOVA problem for DNA microarray data.
\end{abstract}


\end{frontmatter}


\section{Introduction}
In this study, we examine high-dimensional tests for a
bilateral linear hypothesis on the mean matrix in the generalized multivariate analysis of variance (GMANOVA) model. After establishing some prefatory notations, we focus on a more precise problem statement.    
Let $N_i$ independent $p$-dimensional observation vectors, $\bs{x}_{1}^{(i)}$, \ldots, $\bs{x}_{N_i}^{(i)}$, be drawn from the $i$-th population, where $i \in \llbracket g \rrbracket=\{1,\ldots,g\}$ and $g$ denotes the number of underlying  populations. 
For the complete observation matrix, defined as $\bs{X}=(\bs{x}_{1}^{(1)},\ldots,\bs{x}_{N_1}^{(1)},\bs{x}_{1}^{(2)},\ldots,\bs{x}_{N_2}^{(2)},\ldots,\bs{x}_{1}^{(g)},\ldots,\bs{x}_{N_g}^{(g)})'$, we assume the following generalized multivariate linear model:
\begin{equation}
\bs{X} = \bs{A}\bs{\Theta}\bs{B}'+\boldsymbol{\mathcal{E}}, 
\label{eq:generalizedmultivariatelinearmodel}
\end{equation}
where 
$\bs{A}$ is a given $N \times k$ matrix with rank $k$, 
known as a between-group design matrix, $N=N_1+\cdots+N_g$; 
$\bs{\Theta}$ is a $k \times q$ unknown mean parameter matrix; 
$\bs{B}$ is a given $p \times q$ matrix with rank $q$, 
known as a within-design matrix; and
$\boldsymbol{\mathcal{E}} 
=(\bs{\varepsilon}_{1},\ldots,\bs{\varepsilon}_{N})'$
is an $N \times p$ error matrix with mean $\bs{O}$.  
For details of the model $(\ref{eq:generalizedmultivariatelinearmodel})$, please refer to Rosen \cite{Rosen2018}. 
Our primary objective is to develop a test procedure for the matrix parameter $\bs{\Theta}$ by relaxing the commonly adopted linear model assumptions, such as normality and homoscedasticity. 
We assume that 
\[
\mathrm{Var} \left(
\bs{\varepsilon}_{
\scalebox{0.5}{$
\bar{N}_i+j$}
}
\right)
= \bs{\Sigma}_i, \quad 
j \in 
\llbracket N_i \rrbracket, 
\quad i \in 
\llbracket g \rrbracket, 
\]
where $\bs{\Sigma}_i$ denotes a positive-definite $p\times p$ symmetric matrix and   
$\bar{N}_i=N_0+N_1+\cdots+N_{i-1}$ with $N_0=0$, and the distribution of $\bs{X}$ may be non-normal; in addition, $\bs{\Sigma}_1,\ldots,\bs{\Sigma}_g$ may not be equal and $N_1,\ldots,N_{g}$ are allowed not to be equal, thereby implying a heteroscedastic and unbalanced data design.  
Furthermore, in our setup, $p$ is much larger than $N_i$.  

This paper proposes testing statistics for a bilateral linear hypothesis on the mean matrix, which is given as follows: 
\begin{equation}
H_0: \bs{L} \bs{\Theta} \bs{R}' = \bs{O} \quad 
\mathrm{vs.} \quad 
H_1:\bs{L} \bs{\Theta} \bs{R}' \ne \bs{O}, 
\label{eq:hypothesis}
\end{equation}
where $\bs{L}:\ell \times k$ and $\bs{R}:r \times q$ are known matrices with ranks $\ell$ and $r$, respectively. 
The proposed tests are constructed to be well-defined for a high-dimensional GMANOVA model. 
We assume that $\ell$ and $r$ are fixed, even as the observation dimension $p$ and sample size $N_{i}$ tend toward infinity. 

Owing to the general formulation and with various options for $\bs{L}$ and $\bs{R}$, the testing problem (\ref{eq:hypothesis}) incorporates numerous hypotheses of interest. For example, the test for a linear hypothesis on the mean matrix, particularly for testing the homogeneity of the means, is a special case. 

The theory of multivariate inference proposes several solutions to this problem for the classical case of $p<N_{i}$, particularly assuming normality and homoscedasticity. 
The hypotheses of the mean in a multivariate linear model are usually tested based on the likelihood ratio (LR) criterion. 
An extensive overview of the results for a large $N_{i}$ and fixed $p$, along with the related results, is provided in classical multivariate analysis literature; see, e.g., Muirhead \cite{Muirhead1982}, Srivastava \cite{Srivastava2002}, Anderson \cite{Anderson2003}, and Fujikoshi et al. \cite{Fujikoshi2010}. 

The classical methods for a one- and higher-way MANOVA collapse when $p >N_{i}$, mainly from the singularity of the empirical covariance matrix involved, and thus, need to be modified. 
Several suggestions have been recently provided in the literature regarding the modification of the classical MANOVA tests for high-dimensional data. 
While most of these modifications relax only the normality assumption, such as those developed by Srivastava and Kubokawa \cite{Srivastava2013} and Yamada and Himeno \cite{Yamada2015}, there are more flexible approaches offering a completely nonparametric method to the problem, such as those developed by Ghosh and Biswas \cite{Ghosh2016}, Wang et al. \cite{Wang2015}, and Wang et al. \cite{Wang2019}. 

Similarly, the results have been reported under homoscedasticity, i.e., $\bs{\Sigma}_i =\bs{\Sigma}$,  $\forall  i \in \llbracket g \rrbracket$. 
For example, Yamada and Sakurai \cite{Yamada2012} compared the powers of three classical tests under an asymptotic framework in which $p$ and $N$ both increase such that $p/N$ converges to a positive constant. 
Srivastava and Singull \cite{Srivastava2017} provided a test procedure whose testing statistic was based on a modified maximum likelihood estimator of the mean matrix; 
Jana et al. \cite{Jana2017} proposed the test statistic by replacing $\bs{\Sigma}^{-1}$ in the LR statistic derived under the condition that $\bs{\Sigma}$ is known with the Moore-Penrose inverse of the unbiased estimator of $\bs{\Sigma}$. 
All these results were derived for a homoscedastic GMANOVA under normality.  

As another prominent issue, despite demonstrating several important results, most of the modifications of the classical theory have focused on some specific, high-dimensional testing problems, such as those proposed by Cai et al. \cite{Cai2014a}, Cai and Xia \cite{Cai2014b}, and Chen et al. \cite{Chen2019}, who restricted their attention to two-sample tests; 
Zhou et al. \cite{Zhou2017}, who developed a test procedure for 
a hypothesis regarding linear combinations of the mean vectors; 
and Zhou et al. \cite{Zhou2020}, who proposed a test based on the $\boldsymbol{L^{2}}$-norm for a high-dimensional two-way MANOVA. 

The present study, however, aligns with a different approach. 
Instead of exploring a specific testing problem, we aim to develop a unified methodology that encompasses a wide spectrum of high-dimensional testing problems ranging from a one-way MANOVA to tests for parallelism in a profile analysis. 
The construction adopted for the unified theory is based on the bias-corrected estimator of the Frobenius norm of $\boldsymbol{\mathfrak{A}}^{1/2}\bs{L} \bs{\Theta} \bs{R}'\boldsymbol{\mathfrak{B}}^{1/2}$, where $\boldsymbol{\mathfrak{A}}$ and $\boldsymbol{\mathfrak{B}}$ are positive-definite symmetric matrices. 
The main distinguishing feature of the proposed methodology is that we simultaneously relax the classical linear model assumptions, such as normality, homoscedasticity, and low dimensionality, for all cases of a high-dimensional GMANOVA. 
In fact, our proposed tests can be applied for a distribution family that contains normal. 
The asymptotic framework of interest is one where $p$ and $N_{i}$ both increase, while $p/N_{i}$ can converge to any arbitrarily large but bounded positive value. 
The asymptotic theory is developed under general assumptions that include an elliptical distribution and a family of distribution-specified conditions, such as those applied by Bai and Saranadasa \cite{Bai1996} and Chen and Qin \cite{Chen2010a}. 

The remainder of this paper is organized as follows. In Section \ref{sec:Testing statistic and its variance}, we construct our class of statistics to test $H_{0}$ in (\ref{eq:hypothesis}). 
In Section \ref{sec:Asymptotic distribution}, we show that the asymptotic distributions of the proposed test statistics become normal under the null and a specified alternative hypotheses. 
The proofs of the main results are provided in Appendix A, and additional theoretical results are given in the online supplementary material. 
In the online supplementary material, specific results contained in the proposed class of tests are reviewed, and the results of numerical evaluations conducted through various finite-sample simulation scenarios, along with an example using real data are also reported.  

Hereafter, ``$\indep$'' denotes the independence,  
$\Phi(.)$ denotes the cumulative distribution function of the standard normal distribution, 
$\Phi^{-1}(.)$ denotes the inverse function of $\Phi(.)$, 
``$\overset{P}{\to}$'' represents the convergence in probability, 
and ``$\overset{\mathcal{D}}{\to}$'' denotes the convergence in distribution. 
In addition, we define ``$\asymp$'' as the asymptotic equivalence, as follows: 
\[
A \asymp B \overset{\mathrm{def}}{\iff} 
\frac{A}{B} = O(1) \quad \mathrm{and} \quad 
\frac{B}{A} = O(1). 
\]

\section{Testing statistic and its variance}
\label{sec:Testing statistic and its variance}

Before proposing the test statistics, we present a trivial, yet essential, aspect. 
\begin{lem}
\label{lem:equivhypothesis}
Testing the hypothesis $(\ref{eq:hypothesis})$ is equivalent to testing 
\begin{equation}
H_0: \tr(\boldsymbol{\mathfrak{B}}^{1/2} \bs{R} \bs{\Theta}' \bs{L}' 
\boldsymbol{\mathfrak{A}}
\bs{L}\bs{\Theta}\bs{R}' \boldsymbol{\mathfrak{B}}^{1/2}
)=0 
\quad \mathrm{vs.} \quad 
H_1:\tr(\boldsymbol{\mathfrak{B}}^{1/2} \bs{R} \bs{\Theta}' \bs{L}' 
\boldsymbol{\mathfrak{A}}
\bs{L}\bs{\Theta}\bs{R}' \boldsymbol{\mathfrak{B}}^{1/2}
) > 0, 
\label{eq:equivhypothesis}
\end{equation}
where $\boldsymbol{\mathfrak{A}}$ and 
$\boldsymbol{\mathfrak{B}}$ are positive-definite symmetric matrices. 
\end{lem}

Note that 
$\boldsymbol{\mathfrak{A}}$ and 
$\boldsymbol{\mathfrak{B}}$ in Lemma \ref{lem:equivhypothesis} 
are arbitrary; therefore, we set 
\begin{align*}
\boldsymbol{\mathfrak{A}} &= \{ \bs{L}(\bs{A}'\bs{A})^{-1}\bs{L}' \}^{-1}, \quad 
\boldsymbol{\mathfrak{B}} = 
\{ \bs{R}(\bs{B}'\bs{B})^{-1}\bs{R}' \}^{-1}. 
\end{align*}
Then, the expression in (\ref{eq:equivhypothesis}) is described as follows: 
\begin{align*}
\mathcal{Q}&=
\tr(\boldsymbol{\mathfrak{B}}^{1/2} \bs{R} \bs{\Theta}' \bs{L}' 
\boldsymbol{\mathfrak{A}}
\bs{L}\bs{\Theta}\bs{R}' \boldsymbol{\mathfrak{B}}^{1/2}
)\\
&=
\tr(
\{ \bs{R}(\bs{B}'\bs{B})^{-1}\bs{R}' \}^{-1/2}
\bs{R} \bs{\Theta}' \bs{L}' 
\{ \bs{L}(\bs{A}'\bs{A})^{-1}\bs{L}' \}^{-1}
\bs{L}\bs{\Theta}\bs{R}' 
\{ \bs{R}(\bs{B}'\bs{B})^{-1}\bs{R}' \}^{-1/2}
). 
\end{align*}
We prepare an unbiased estimator of $\mathcal{Q}$ to define the testing statistic. 
A natural estimator of $\mathcal{Q}$ is given as follows:
\[
\widehat{\mathcal{Q}}= \tr 
(
\{ \bs{R}(\bs{B}'\bs{B})^{-1}\bs{R}' \}^{-1/2}
\bs{R} \hat{\bs{\Theta}}' 
\bs{L}' 
\{ \bs{L}(\bs{A}'\bs{A})^{-1}\bs{L}' \}^{-1}
\bs{L} \hat{\bs{\Theta}} 
\bs{R}' 
\{ \bs{R}(\bs{B}'\bs{B})^{-1}\bs{R}' \}^{-1/2}
), 
\]
where 
$\hat{\bs{\Theta}} = (\bs{A}'\bs{A})^{-1}\bs{A}'\bs{X}\bs{B}(\bs{B}'\bs{B})^{-1}$, which is an unbiased estimator of $\bs{\Theta}$. 
Through 
\begin{align*}
\boldsymbol{\mathcal{P}} &= 
\{ \bs{R}(\bs{B}'\bs{B})^{-1}\bs{R}' \}^{-1/2} 
\bs{R} (\bs{B}'\bs{B})^{-1} \bs{B}',\\
\bs{\Pi}_H &= 
\bs{A}(\bs{A}'\bs{A})^{-1} \bs{L}' 
\{ \bs{L}(\bs{A}'\bs{A})^{-1}\bs{L}' \}^{-1}
\bs{L} (\bs{A}'\bs{A})^{-1} \bs{A}',
\end{align*}
we can simply denote $\widehat{\mathcal{Q}}=\tr (\boldsymbol{\mathcal{P}} \bs{X}' \bs{\Pi}_{\mathrm{H}} \bs{X} \boldsymbol{\mathcal{P}}')$.  
Note that $\boldsymbol{\mathcal{P}}$ is a
$r \times p$ matrix with rank $r$, such that  
$\boldsymbol{\mathcal{P}}'\boldsymbol{\mathcal{P}}$ 
is a projection matrix and $\bs{\Pi}_H$ is an $N \times N$ projection matrix with rank $\ell$. It can then be stated that 
\begin{align}
\widehat{\mathcal{Q}}
&= \tr (\boldsymbol{\mathcal{P}} \bs{B} \bs{\Theta}' \bs{A}' 
\bs{\Pi}_{\mathrm{H}} 
\bs{A} \bs{\Theta} \bs{B}' \boldsymbol{\mathcal{P}}')
+ 2 
\tr (\boldsymbol{\mathcal{P}} \bs{B} \bs{\Theta}' \bs{A}'
\bs{\Pi}_{\mathrm{H}} 
\boldsymbol{\mathcal{E}}' \boldsymbol{\mathcal{P}}')
\nonumber \\
& \quad +\sum_{i,j}^N h_{ij} 
\tr 
(
\boldsymbol{\mathcal{P}} 
\bs{\varepsilon}_i 
\bs{\varepsilon}_j' 
\boldsymbol{\mathcal{P}}'  
)
+\sum_{i=1}^N h_{ii} 
\bs{\varepsilon}_i' 
\boldsymbol{\mathcal{P}}'  
\boldsymbol{\mathcal{P}} 
\bs{\varepsilon}_i, 
\label{eq:expressionhatQ}
\end{align}
where $(h_{ij})=\bs{\Pi}_{\mathrm{H}}$ and $\sum_{i_1,\ldots,i_k}^N$ denote the sum over all different indices $\{i_1,\ldots,i_k\}$; for example, $\sum_{i,j,k}^N = \sum_{i=1}^N \sum_{\substack{j=1 \\ j \ne i}}^N\sum_{\substack{k=1 \\ k \ne i, k \ne j}}^N. $
It is clear that $\sum_{i=1}^N h_{ii} \bs{\varepsilon}_i' 
\boldsymbol{\mathcal{P}}'\boldsymbol{\mathcal{P}} 
\bs{\varepsilon}_i$ on the right side of equality $(\ref{eq:expressionhatQ})$ is an unnecessary term for an unbiased estimate of $\widehat{\mathcal{Q}}$. 
To omit this term, we prepare an adjustment that is derived 
from the assumption that the linear system given below has the following solution: 
\begin{equation}
[(\bs{I}_N-\bs{\Pi}_{\boldsymbol{A}}) \odot (\bs{I}_N-\bs{\Pi}_{\boldsymbol{A}})]
\begin{pmatrix} d_1 \\ \vdots \\ d_N \end{pmatrix}
= 
\begin{pmatrix} h_{11} \\ \vdots \\ h_{NN} \end{pmatrix}, 
\label{eq:linearequation}
\end{equation}
where 
the notation ``$\odot$'' denotes the Hadamard product of 
the matrices. 
By letting $\bs{D}=\mathrm{diag}(d_1,\ldots,d_N)$, 
the following formula can be obtained: 
\begin{align*}
&\tr 
\left(
\boldsymbol{\mathcal{P}} \bs{X}' 
(\bs{I}_N-\bs{\Pi}_{\boldsymbol{A}}) \bs{D} (\bs{I}_N-\bs{\Pi}_{\boldsymbol{A}}) \bs{X} \boldsymbol{\mathcal{P}}'
\right)\\
&=
\tr 
\left(
(\bs{I}_N-\bs{\Pi}_{\boldsymbol{A}}) \bs{D} (\bs{I}_N-\bs{\Pi}_{\boldsymbol{A}}) 
\boldsymbol{\mathcal{E}} \boldsymbol{\mathcal{P}}'
\boldsymbol{\mathcal{P}} \boldsymbol{\mathcal{E}}'
\right)
\\
&=
\sum_{i=1}^N 
\left\{
\sum_{k=1}^N d_k (\delta_{ik}-p_{ik})^2
\right\}
\bs{\varepsilon}_i' 
\boldsymbol{\mathcal{P}}' 
\boldsymbol{\mathcal{P}} 
\bs{\varepsilon}_i 
+
\sum_{i,j}^N 
\left\{
\sum_{k=1}^N d_k(\delta_{ik}-p_{ik})
(\delta_{jk}-p_{jk})
\right\}
\bs{\varepsilon}_i' 
\boldsymbol{\mathcal{P}}' 
\boldsymbol{\mathcal{P}} 
\bs{\varepsilon}_j\\
&=
\sum_{i=1}^N 
h_{ii}
\bs{\varepsilon}_i' 
\boldsymbol{\mathcal{P}}' 
\boldsymbol{\mathcal{P}} 
\bs{\varepsilon}_i 
+
\sum_{i,j}^N 
\left\{
\sum_{k=1}^N d_k(\delta_{ik}-p_{ik})
(\delta_{jk}-p_{jk})
\right\}
\tr (
\boldsymbol{\mathcal{P}} 
\bs{\varepsilon}_i
\bs{\varepsilon}_j'
\boldsymbol{\mathcal{P}}' 
), 
\end{align*}
where $\delta_{ij}$ is the Kronecker delta and $(p_{ij})=\bs{\Pi}_{\boldsymbol{A}}$. 
Then, defining
\[
T=\widehat{\mathcal{Q}}-\tr 
\left(
\boldsymbol{\mathcal{P}} \bs{X}' 
(\bs{I}_N-\bs{\Pi}_{\boldsymbol{A}}) \bs{D} (\bs{I}_N-\bs{\Pi}_{\boldsymbol{A}}) \bs{X} \boldsymbol{\mathcal{P}}'
\right),  
\]
the following equalities hold. 
\begin{align}
T &= 
\tr (\boldsymbol{\mathcal{P}} \bs{B} \bs{\Theta}' \bs{A}' 
\bs{\Pi}_{\mathrm{H}}  
\bs{A} \bs{\Theta} \bs{B}' \boldsymbol{\mathcal{P}}')
+ 2 
\tr (\boldsymbol{\mathcal{P}} \bs{B} \bs{\Theta}' \bs{A}'
\bs{\Pi}_{\mathrm{H}} 
\boldsymbol{\mathcal{E}}' \boldsymbol{\mathcal{P}}')
+\sum_{i,j}^N \omega_{ij} 
\tr 
(
\boldsymbol{\mathcal{P}} 
\bs{\varepsilon}_i 
\bs{\varepsilon}_j' 
\boldsymbol{\mathcal{P}}' 
)
\nonumber\\
&=
\tr (\boldsymbol{\mathcal{P}} \bs{B} \bs{\Theta}' \bs{A}' 
\bs{\Omega}
\bs{A} \bs{\Theta} \bs{B}' \boldsymbol{\mathcal{P}}')
+ 2 
\tr (\boldsymbol{\mathcal{P}} \bs{B} \bs{\Theta}' \bs{A}'
\bs{\Omega}
\boldsymbol{\mathcal{E}}' \boldsymbol{\mathcal{P}}')
+
\tr ( \boldsymbol{\mathcal{P}} \boldsymbol{\mathcal{E}}' \bs{\Omega} \boldsymbol{\mathcal{E}} \boldsymbol{\mathcal{P}}' )
\nonumber\\
&= \tr ( \boldsymbol{\mathcal{P}} \bs{X}' \bs{\Omega} \bs{X} \boldsymbol{\mathcal{P}}' ), 
\label{eq:statT1}
\end{align}
where $\omega_{ij}=h_{ij}-\sum_{k=1}^N d_k(\delta_{ik}-p_{ik})
(\delta_{jk}-p_{jk})$, which yields $
(\omega_{ij})=
\bs{\Omega} = \bs{\Pi}_{\mathrm{H}} - 
(\bs{I}_N-\bs{\Pi}_{\boldsymbol{A}}) \bs{D} 
(\bs{I}_N-\bs{\Pi}_{\boldsymbol{A}})$, and 
$\bs{\Omega}$ is a symmetric matrix
such that each of its diagonal elements is zero. 
The second equality in $(\ref{eq:statT1})$ 
is derived from the following:
\begin{align*}
& \tr \left(
\boldsymbol{\mathcal{P}} \bs{B} \bs{\Theta}' \bs{A}' 
(\bs{I}_N-\bs{\Pi}_{\boldsymbol{A}})
\bs{D}
(\bs{I}_N-\bs{\Pi}_{\boldsymbol{A}})
\boldsymbol{\mathcal{E}}'
\boldsymbol{\mathcal{P}}'
\right) = 0, \\
& \tr \left(
\boldsymbol{\mathcal{P}} \bs{B} \bs{\Theta}' \bs{A}' 
(\bs{I}_N-\bs{\Pi}_{\boldsymbol{A}})
\bs{D}
(\bs{I}_N-\bs{\Pi}_{\boldsymbol{A}})
\bs{A} \bs{\Theta} \bs{B}' 
\boldsymbol{\mathcal{P}}
\right) = 0. 
\end{align*}
The expected value of $T$ is then obtained as follows: 
\begin{align*}
E[T] &= 
\tr (\boldsymbol{\mathcal{P}} \bs{B} \bs{\Theta}' \bs{A}' 
\bs{\Omega}
\bs{A} \bs{\Theta} \bs{B}' \boldsymbol{\mathcal{P}}')
+ 2 E[
\tr (\boldsymbol{\mathcal{P}} \bs{B} \bs{\Theta}' \bs{A}' 
\bs{\Omega}
\boldsymbol{\mathcal{E}}
\boldsymbol{\mathcal{P}}')
]
+ E[ \tr ( 
\bs{\Omega} \boldsymbol{\mathcal{E}} \boldsymbol{\mathcal{P}}'
\boldsymbol{\mathcal{P}} \boldsymbol{\mathcal{E}}' 
)] = \mathcal{Q}, 
\end{align*}
i.e., $T$ is an unbiased estimator of $\mathcal{Q}$. From Lemma \ref{lem:equivhypothesis}, 
the rejection of the null hypothesis $H_0$ results from the 
evidence indicating that 
the unbiased estimator of $\mathcal{Q}$ is significantly larger than $0$; 
hence, we propose the testing statistic as $T$. 

The variance of $T$ is described as follows: 
\[
\sigma^2=\mathrm{Var}(T)
=\sigma_0^2+\mathrm{Var}(
\tr \bs{\Omega} (
\bs{A}\bs{\Theta}\bs{B}'
\boldsymbol{\mathcal{P}}'
\boldsymbol{\mathcal{P}} 
\boldsymbol{\mathcal{E}}
+
\boldsymbol{\mathcal{E}}
\boldsymbol{\mathcal{P}}'
\boldsymbol{\mathcal{P}} 
\bs{B}\bs{\Theta}'\bs{A}'
)
), 
\]
where 
\begin{align*}
&\sigma_0^2=\mathrm{Var}(
\tr (\bs{\Omega} 
\boldsymbol{\mathcal{E}}
\boldsymbol{\mathcal{P}}'
\boldsymbol{\mathcal{P}} 
\boldsymbol{\mathcal{E}}')
)
=2 \sum_{i,j}^N 
\omega_{ij}^2 
E[
(\bs{\varepsilon}_i'
\boldsymbol{\mathcal{P}}'
\boldsymbol{\mathcal{P}} 
\bs{\varepsilon}_j
)^2],\\
&\mathrm{Var}(
\tr (
\bs{\Omega} (
\bs{A}\bs{\Theta}\bs{B}'
\boldsymbol{\mathcal{P}}'
\boldsymbol{\mathcal{P}} 
\boldsymbol{\mathcal{E}}'
+
\boldsymbol{\mathcal{E}}
\boldsymbol{\mathcal{P}}'
\boldsymbol{\mathcal{P}} 
\bs{B}\bs{\Theta}'\bs{A}'
))
)
=
4 
\sum_{i=1}^N 
\bs{m}_{(i)}'
E[
\bs{\varepsilon}_i
\bs{\varepsilon}_i']
\bs{m}_{(i)}, \\
& \bs{m}_{(i)} 
= \sum_{j=1}^N \omega_{ij} \bs{m}_j, \quad 
\bs{m}_j = 
\boldsymbol{\mathcal{P}}'
\boldsymbol{\mathcal{P}}
\bs{B}\bs{\Theta}' \bs{a}_j, \quad 
(\bs{a}_1,\ldots,\bs{a}_N) = \bs{A}'. 
\end{align*}
Note that $\sigma^2=\sigma_0^2$ when $H_0$ is true. 
Defining
\begin{align}
\bs{V} &=
\begin{pmatrix}
a_{1,2}
\bs{1}_{N_1} \bs{1}_{N_1}' 
& 
b_{12}
\bs{1}_{N_1} \bs{1}_{N_2}' 
& \cdots 
& 
b_{1g} 
\bs{1}_{N_1} \bs{1}_{N_g}' \\
b_{21}
\bs{1}_{N_2} \bs{1}_{N_1}' 
& 
a_{2,2} 
\bs{1}_{N_2} \bs{1}_{N_2}' 
& \cdots 
& 
b_{2g} 
\bs{1}_{N_2} \bs{1}_{N_g}' \\
\vdots & \vdots & \ddots & \vdots \\
b_{g1}
\bs{1}_{N_g} \bs{1}_{N_1}' 
& 
b_{g2}
\bs{1}_{N_g} \bs{1}_{N_2}' 
& \cdots 
& 
a_{g,2}
\bs{1}_{N_g} \bs{1}_{N_g}' 
\end{pmatrix}, 
\label{eq:defV}
\end{align}
where
\[
a_{i,j}=\tr (
\bs{\Psi}_i^j
),
\quad 
b_{ij} = b_{ji} = 
\tr 
(
\bs{\Psi}_i \bs{\Psi}_j
)
\]
and $ \bs{\Psi}_i=\boldsymbol{\mathcal{P}} \bs{\Sigma}_i \boldsymbol{\mathcal{P}}'$, 
the variance can be described as $\sigma_0^2 = 2 \tr ((\bs{\Omega} \odot \bs{\Omega}) \bs{V}).$

\section{Asymptotic distribution}
\label{sec:Asymptotic distribution}

This section presents an 
asymptotic distribution for testing statistic $T$ under the following assumptions: 
\begin{enumerate}
\item[A1:] $
\displaystyle 
\limsup_{\min\{ N_1,\ldots,N_g\} \to \infty} 
\rho_N < \infty$, 
where 
$
\displaystyle 
\rho_N=
\frac{
\max \{
\omega_{ij}^2~|~
i,j \in \llbracket N \rrbracket, i < j
\}
}{
\min \{
\omega_{ij}^2~|~
i,j \in \llbracket N \rrbracket, i < j, \omega_{ij} \ne 0
\}
}
$. 
\item[A2:] 
$ 
\displaystyle 
\lim_{p \to \infty} 
\max_{
\substack{(i_k i_{\ell}) \in \mathfrak{I} \\ 
k,\ell \in \{1,2,3,4\}, k \ne \ell
}}
\psi_{i_1 i_2 i_3 i_4 }
\Bigg{/}
\left(
\sum_{(i,j) \in \mathfrak{I}} b_{ij}
\right)^2
=0$, 
where $\psi_{ijk \ell} = \tr( \bs{\Psi}_i \bs{\Psi}_j \bs{\Psi}_k \bs{\Psi}_{\ell} )$; 
$\sum_{(i,j) \in \mathfrak{I}} b_{ij}$ denotes the sum 
of $b_{ij}$ for indices $i,j \in \llbracket g \rrbracket$, such that $(i,j) \in \mathfrak{I}$, 
$\mathfrak{I} = 
\{(i,j)~|~i,j \in \llbracket g \rrbracket, 
\bs{1}_{N_i}'
(\bs{\Omega}_{ij} \odot \bs{\Omega}_{ij}) \bs{1}_{N_j}>0\}$; 
and $\bs{\Omega}_{ij}$ is the $(i,j)$-th block in $\bs{\Omega}$, 
which corresponds to that of $\bs{V}$ in $(\ref{eq:defV})$. 
\item[A3:] 
$ \displaystyle 
\lim_{\min \{N_1,\ldots,N_g,p\} \to \infty}
\frac{
\sum_{i=1}^g 
\sum_{j=1}^{N_i} 
(\bs{m}_{
(\scalebox{0.5}{$
\bar{N}_i+j$}
)}'
\bs{\Sigma}_i
\bs{m}_{
(\scalebox{0.5}{$
\bar{N}_i+j$}
)})^2
}{
\mathcal{M}+(1-\mathcal{M})
\left\{
\sum_{i=1}^g 
\sum_{j=1}^{N_i} 
\bs{m}_{
(\scalebox{0.5}{$
\bar{N}_i+j$}
)}'
\bs{\Sigma}_i
\bs{m}_{
(\scalebox{0.5}{$
\bar{N}_i+j$}
)}
\right\}^2
} 
=0$, where 
\[
\mathcal{M}
=\left\{
\begin{array}{ll}
1, & \mbox{$
\bs{m}_{(i)}' \bs{m}_{(i)}=0$ for $ i \in 
\llbracket N \rrbracket$}, \\
0 ,& \mbox{otherwise}. 
\end{array} 
\right. 
\]
\end{enumerate}
For the two sample problems presented in online supplementary material, $\boldsymbol{\mathcal{P}}$ becomes the identity matrix and $\mathcal{I}=\{1,2\} \times \{1,2\}$; thus, $\mathrm{A2}$ is identical to the condition described by Chen and Qin \cite{Chen2010a}. 
From this perspective, $\mathrm{A2}$ can be viewed as a modification of their condition. 
Note also that the assumption A3 is always satisfied under $H_0$. 

We consider the distribution of $T$ under the following model: 
\begin{equation}
\bs{\varepsilon}_{\bar{N}_i+j} 
=\bs{\Sigma}_i^{1/2} \bs{z}_{\bar{N}_i+j}, \quad 
j \in 
\llbracket N_i \rrbracket, 
\quad i \in 
\llbracket g \rrbracket, 
\label{eq:errortermmod}
\end{equation}
where $\bs{z}_i,\ldots,\bs{z}_N$ are independently and identically distributed (i.i.d.) as a $p$-dimensional distribution $F$, 
with mean $\bs{0}$ and covariance matrix $\bs{I}_p$. 
In addition, we assume the following. 
For $\bs{z}=(z_1,\ldots,z_p)' \sim F$, 
\begin{enumerate}
\item[D1:] $\limsup_{p \to \infty} \max \{ E[z_i^4]~|~ 
i \in \llbracket p \rrbracket \} < \infty$. 
\item[D2:] $E[\prod_{j=1}^{p} z_{j}^{\nu_j}]=0$ 
when there is at least one $\nu_i=1$  
whenever $\nu_1+\cdots+\nu_p=4$. 
\end{enumerate}

Notably, $\mathrm{D2}$ is milder than the condition used by Bai and Saranadasa \cite{Bai1996}, without assuming that $E[z_i^2z_j^2]=1$ for $i,j \in \llbracket g \rrbracket$ with $i \ne j$; 
moreover, the set of $\mathrm{D1}$ and $\mathrm{D2}$ is milder than the condition used by Chen and Qin \cite{Chen2010a}, in which $E[z_i^4]=3+\gamma$ and $E[\prod_{i=1}^q z_{\ell_i}^{\nu_i}]=\prod_{i=1}^q E[z_{\ell_i}^{\nu_i}]$, where $q$ is a positive integer such that $\sum_{i=1}^q \nu_i \leq 8$ and $\ell_1 \ne \cdots \ne \ell_q$.
In addition, conditions $\mathrm{D1}$ and $\mathrm{D2}$ are valid when $F$ is an elliptical distribution with mean $\bs{0}$, covariance matrix $\bs{I}_p$, and a finite fourth moment. 
Thus, our assumption covers a wide range of multivariate distributions, including $N_p(\bs{0}, \bs{I}_p)$.  

\begin{remark}
Bai et al. \cite{Bai2018} derived the asymptotic joint distributions 
of the eigenvalues in the MANOVA model as $N,p \to \infty$ 
under the condition that $p/N \to c \in (0,1)$, based on the random matrix theory. 
For the error vector $\bs{z}=(z_1,\ldots,z_p)$, 
the authors assumed that $z_1,\ldots,z_p$ are i.i.d.; 
$E[z_i]=0$, $E[z_i^2]=1$, and $E[z_i^4]<\infty$. 
Because their assumption implies $\mathrm{D1}$ and $\mathrm{D2}$, 
our distribution family also includes their assumptions. 
\end{remark}

\begin{remark}
Yamada and Himeno \cite{Yamada2015} also used  an assumption to specify the distribution family of $F$ including multidimensional normal distribution. 
Note that their assumption is satisfied when $\mathrm{A1}$-$\mathrm{A3}$, 
$\mathrm{D1}$, and $\mathrm{D2}$ hold, and thus, is milder than our own. 
However, we do not use Yamada and Himeno \cite{Yamada2015}'s assumption because it is in a complex form. 
\end{remark}

\begin{thm}
\label{thm:asympdis}
Suppose that the linear system in $(\ref{eq:linearequation})$ 
has a solution, and that 
$\mathrm{A1}$-$\mathrm{A3}$, $\mathrm{D1}$, and $\mathrm{D2}$ hold. 
Then, 
\[
\frac{
T-Q
}{
\sqrt{\sigma^2}
} \overset{\mathcal{D}}{\to} N(0,1)
\]
under the asymptotic framework that $N_i \asymp N_j$ for $i, j \in \llbracket g \rrbracket$ as 
$\min\{N_1,\ldots,N_g,p\}$ tends toward infinity. 
\end{thm}
\noindent
The proof is given in the Appendix. 


\subsection{Asymptotic null distribution and proposed testing criterion}
In this subsection, we present an asymptotic null distribution of $T$. 
From Theorem \ref{thm:asympdis}, under the condition that the null hypothesis 
$H_0$ is true, 
\[
\frac{T}{\sqrt{\sigma_0^2}} \overset{\mathcal{D}}{\to} N(0,1).  
\]
To analyze the real data, $\sigma_0^2$ must be estimated. 
Partition 
\[
\bs{X}=(\bs{X}_1',\ldots,\bs{X}_g)', \quad 
\bs{A}=(\bs{A}_1',\ldots,\bs{A}_g')', 
\]
where $\bs{X}_i=(\bs{x}_{1}^{(i)},\ldots,\bs{x}_{N_i}^{(i)})':N_i \times p$ 
and $\bs{A}_i:N_i \times k$. 
Define 
\begin{align*}
\bs{S}_i &= 
\frac{1}{N_i-k_i} 
\sum_{j=1}^{N_i} 
\boldsymbol{\mathcal{P}}
(\bs{x}_j^{(i)}-\hat{\bs{x}}_j^{(i)})
(\bs{x}_j^{(i)}-\hat{\bs{x}}_j^{(i)})'
\boldsymbol{\mathcal{P}}'
=
\frac{1}{N_i-k_i} 
\boldsymbol{\mathcal{P}}
\bs{X}_i' (\bs{I}_{N_i}-\bs{\Pi}_{\boldsymbol{A}_i}) \bs{X}_i \boldsymbol{\mathcal{P}}'
,\\
Q_i &= \frac{1}{N_i-k_i} \sum_{j=1}^{N_i} 
\left\{
(\bs{x}_j^{(i)}-\hat{\bs{x}}_j^{(i)})'
\boldsymbol{\mathcal{P}}'
\boldsymbol{\mathcal{P}}
(\bs{x}_j^{(i)}-\hat{\bs{x}}_j^{(i)})
\right\}^2,
\end{align*}
where $\bs{\Pi}_{\boldsymbol{A}_i}=\bs{A}_i(\bs{A}_i'\bs{A}_i)^{+}\bs{A}_i'$, 
$k_i = \mathrm{rank}(\bs{A}_i)$, 
\[
(\hat{\bs{x}}_1^{(i)},\ldots,\hat{\bs{x}}_{N_i}^{(i)})'
=\bs{\Pi}_{\boldsymbol{A}_i}
(\bs{x}_1^{(i)},\ldots,\bs{x}_{N_i}^{(i)})'
=\bs{\Pi}_{\boldsymbol{A}_i} \bs{X}_i. 
\]
Here, $\bs{A}^+$ is the Moore-Penrose inverse of the matrix $\bs{A}$. 
In the following lemma, 
we provide an expression for the unbiased estimator of 
$a_{i,2}$ 
and that of $b_{ij}$. 

\begin{lem}
\label{lem:unbiased estimator}
The unbiased estimators of $a_{i,2}$ and $b_{ij}$ 
are expressed as follows:
\begin{align*}
\hat{a}_{i,2} &= 
\frac{1}{(N_i-k_i) \tau_{i,3}}
\left[
\{(N_i-k_i)^2 \tau_{i,2} - \tau_{i,1}^2\} \tr (\bs{S}_i^2) 
-\{(N_i-k_i) \tau_{i,2} - \tau_{i,1}^2\} (\tr (\bs{S}_i))^2 
\right. \\
& \quad \left. 
-(N_i-k_i-1) \tau_{i,2} Q_i
\right], \\
\hat{b}_{ij} &=  \tr (\bs{S}_i \bs{S}_j), 
\end{align*}
respectively, where 
\begin{align*}
\tau_{i,j} &=
\tr (
\{ (\bs{I}_{N_i}-\bs{\Pi}_{\boldsymbol{A}_i}) \odot  
(\bs{I}_{N_i}-\bs{\Pi}_{\boldsymbol{A}_i}) \}^j), \quad j=1,2, \\
\tau_{i,3} &= \frac{N_i-k_i-1}{(N_i-k_i)^2}
\{ (N_i-k_i)(N_i-k_i+2) \tau_{i,2} - 3 \tau_{i,1}^2 \}. 
\end{align*}
\end{lem}
\noindent
The unbiasedness of $\hat{b}_{ij}$ is immediately followed from the 
fact that $\bs{S}_i$ and $\bs{S}_j$ are independent 
and that $E[\bs{S}_i]=\bs{\Psi}_i$. 
The unbiasedness of $\hat{a}_{i,2}$ is proved in the Supplementary Materials. 

As an example of Lemma \ref{lem:unbiased estimator}, set 
\begin{equation}
\bs{A}=\begin{pmatrix}
\bs{1}_{N_1} & \bs{0} & \cdots & \bs{0} \\
\bs{0} & \bs{1}_{N_2} & \cdots & \bs{0} \\
\vdots & \vdots & \ddots & \vdots \\
\bs{0} & \bs{0} & \cdots & \bs{1}_{N_g}
\end{pmatrix}
=
\mathrm{diag}(\bs{1}_{N_1},\ldots,\bs{1}_{N_g}).  
\label{eq:MANOVAsettingA}
\end{equation}
It holds that 
\begin{align*}
\bs{\Pi}_{\boldsymbol{A}_i}
&=
\frac{1}{N_i} \bs{1}_{N_i} \bs{1}_{N_i}'
=\bs{\Pi}_{N_i}, 
\quad 
i \in \llbracket g \rrbracket, 
\end{align*}
and thus, 
\[
\bs{\Pi}_{\boldsymbol{A}} = 
\bs{A} \diag(N_1^{-1},\ldots,N_g^{-1}) \bs{A}'
=\mathrm{diag}
(\bs{\Pi}_{N_1}
,\bs{\Pi}_{N_2},\ldots,
\bs{\Pi}_{N_g}
). 
\]
From this result, the following equalities hold. 
\begin{align*}
&\tau_{1,i}
=\frac{(N_i-1)^2}{N_i}
, \quad 
\tau_{2,i}=\frac{
(N_i-1)(N_i^2-3N_i+3)
}{
N_i^2
}, \quad 
\tau_{3,i}
=\frac{
(N_i-2)^2(N_i-3)}{
N_i
}, \\
& \hat{\bs{x}}_j^{(i)}
=\bar{\bs{x}}^{(i)}
=\frac{1}{N_i} \sum_{j=1}^{N_i} \bs{x}_j^{(i)}
, \quad 
j \in \llbracket N_i \rrbracket, 
\quad 
i \in \llbracket g \rrbracket. 
\end{align*}
Therefore, $\hat{a}_{i,2}$ in Lemma \ref{lem:unbiased estimator} can be described as 
\begin{equation}
\hat{a}_{i,2}
=
\frac{N_i-1}{N_i(N_i-2)(N_i-3)}
\left\{
(N_i-1)(N_i-2) \tr (\bs{S}_i^2)
+(\tr (\bs{S}_i))^2
-N_i Q_i
\right\}. 
\label{eq:hy2014consisest}
\end{equation}
This result coincides with the expression of Himeno and Yamada \cite{Himeno2014} 
for the unbiased estimator of $\tr (\bs{\Sigma}_i^2)$. 
Note that
the expression $(\ref{eq:hy2014consisest})$ 
is identical to the following expression: 
\[
\hat{a}_{i,2}
= 
\frac{1}{\mathrm{P}_{N_i,4}} 
\sum_{k,\ell,\alpha,\beta}^{N_i} 
\frac{
\{(\bs{x}_k^{(i)}-\bs{x}_{\ell}^{(i)})'(\bs{x}_{\alpha}^{(i)}-\bs{x}_{\beta}^{(i)})\}^2
}{4}, 
\]
where the symbol $\mathrm{P}_{n,k}$ 
denotes the number of such $k$-permutations of $n$, i.e., 
\[
\mathrm{P}_{n,k}
=n(n-1) \cdots (n-k+1).
\]
Thus, we find that $\hat{a}_{i,2}$ always takes a positive value with probability 1. 

\begin{remark}
\label{remark:non-nagative}
For the general structure of $\bs{A}$, $\hat{a}_{i,2}$ does not always take a 
non-negative value. 
\end{remark}

We mention that $\hat{a}_{2, i}$ and $\hat{b}_{ij}$ have rate consistencies, 
which are represented by the following theorem:

\begin{thm}
\label{thm:rate-consistency}
Suppose that $\hat{a}_{2, i}$ and $\hat{b}_{ij}$ are defined as Lemma \ref{lem:unbiased estimator}. 
Then, under the assumptions $\mathrm{A1}$-$\mathrm{A2}$ and $\mathrm{D1}$-$\mathrm{D2}$, $\hat{a}_{i, 2}/a_{i, 2}$ converges in probability to 1 under the asymptotic framework that $p/N_i$ converges to a non-negative constant as $\min\{p, N_i\}$ tends toward  infinity for $i \in \llbracket g \rrbracket$; $\hat{b}_{ij}/b_{ij}$ converges in probability to 1 as $\min\{p, N_i, N_j\}$ tends toward infinity for $i, j \in \llbracket g \rrbracket$ with $i \ne j$. 
\end{thm}


\begin{remark}
Because $\hat{a}_{i,2}$ and $\hat{b}_{ij}$ are location-free estimators, 
their unbiasedness and rate consistency hold under $H_1$.  
\end{remark}

Let $\hat{\sigma}_0^2$ be the estimator of $\sigma_0^2$ obtained by replacing the unknown parameters in $\sigma_0^2$ with their unbiased estimators given in Lemma \ref{lem:unbiased estimator}, i.e., $\hat{\sigma}_0^2=\tr ((\bs{\Omega} \odot \bs{\Omega}) \hat{\bs{V}})$, where $\hat{\bs{V}}$ is defined by $\bs{V}$ through a replacement of the unknown parameters with their unbiased estimators. 
Then, 
\begin{equation}
\frac{
\hat{\sigma}_0^2 
}{
\sigma_0^2
}
\overset{P}{\to} 1 
\label{eq:rateconsistsigma}
\end{equation}
under the asymptotic framework that $\max\{p/N_i : i \in \llbracket g \rrbracket \}$ converges to a non-negative constant as $\min\{p,N_1,\ldots,N_g\}$ tends toward infinity. 

\begin{remark}
As mentioned in Remark \ref{remark:non-nagative}, 
$\hat{a}_{i,2}$ does not always take a 
non-negative value, which indicates that 
$\hat{\sigma}_0^2$ also does not always take a
non-negative value. 
\end{remark}


\begin{thm}
\label{thm:asymptoticnulldistribution}
Suppose that the assumptions in Theorem 
\ref{thm:asympdis} are satisfied. 
Then, under the null hypothesis $H_0$ given in $(\ref{eq:hypothesis})$,  
\[
I(\hat{\sigma}_0^2 > 0)
\frac{T}{\sqrt{\hat{\sigma}_0^2}} 
\overset{\mathcal{D}}{\to} N(0, 1) 
\]
under the asymptotic framework that $\max\{p/N_i : i \in \llbracket g \rrbracket \}$ converges to a non-negative constant as $\min\{p, N_1, \ldots, N_g\}$ tends toward infinity and $N_i \asymp N_j$ for $i, j \in \llbracket g \rrbracket$, where $I(.)$ is the indicator function, and $\hat{\sigma}_0^2$ is defined as $\sigma_0^2$ by replacing $a_{i, 2}$ and $b_{ij}$ with $\hat{a}_{i, 2}$ and $\hat{b}_{ij}$, respectively, as defined in Lemma \ref{lem:unbiased estimator}. 
\end{thm}

We propose the following testing criterion with the significance level $\varepsilon$ for the case in which the linear system in $(\ref{eq:linearequation})$ has a solution.  
If $\tr ((\bs{\Omega} \odot \bs{\Omega}) \hat{\bs{V}}) > 0$, then 
\begin{equation}
\frac{T}{\sqrt{\hat{\sigma}_0^2}}
=\frac{T}{\sqrt{
2 \tr ((\bs{\Omega} \odot \bs{\Omega}) \hat{\bs{V}}) 
}} > \Phi^{-1}(1-\varepsilon) 
\quad 
\Longrightarrow \quad \mbox{Reject the null hypothesis $H_0$}. 
\label{eq:proposedtest}
\end{equation}

\subsection{Asymptotic power for the proposed test}
In this section, the asymptotic power for the proposed test is considered. 
Note that the following equivalence holds. 
\begin{align*}
I(\hat{\sigma}_0^2 > 0)
\frac{T}{\sqrt{\hat{\sigma}_0^2}} > \Phi^{-1}(1-\varepsilon) 
& \iff 
\frac{T - \mathcal{Q}}{
\sqrt{\sigma^2}}
> 
I(\hat{\sigma}_0^2 > 0)
\sqrt{\frac{\hat{\sigma}_0^2}{\sigma^2}}
\Phi^{-1}(1-\varepsilon) 
-\frac{\mathcal{Q}
}{\sqrt{\sigma^2}}. 
\end{align*}
Because the ratio consistency of $\hat{\sigma}_0^2$ given in $(\ref{eq:rateconsistsigma})$ also holds under $H_1$, Theorem \ref{thm:asympdis} yields the asymptotic power of the proposed test, which is described as follows.   
\begin{thm}
\label{thm:asymptoticpower}
Suppose that the assumptions in Theorem 
\ref{thm:asympdis} are satisfied. 
Under the asymptotic framework $\mathrm{A}$ that $\max \{p/N_1, \ldots, p/N_2 \}$ converges to a non-negative constant as $\min\{p, N_1, \ldots, N_g\}$ tends toward infinity and $N_i \asymp N_j$ for $i, j \in \llbracket g \rrbracket$, the asymptotic power becomes 1 if $\mathcal{Q}/\sqrt{\sigma^2} \to \infty$. 
In addition, 
\[
\lim_{\mathrm{A}} 
P \left( 
I(\hat{\sigma}_0^2 > 0)
\frac{T}{\sqrt{\hat{\sigma}_0^2}} > \Phi^{-1}(1-\varepsilon) \right)
=\Phi \left(
\lim_{\mathrm{A}} 
\left( 
-\sqrt{\frac{\sigma_0^2}{\sigma^2}} 
\Phi^{-1}(1-\varepsilon)+
\frac{\mathcal{Q}}{\sqrt{\sigma^2}}
\right)
\right) 
\]
if $\mathcal{Q}/\sqrt{\sigma^2}$ converges to a non-negative constant, where the notation $\lim_{\mathrm{A}}$ denotes the limit under the asymptotic framework $\mathrm{A}$, and $\hat{\sigma}_0^2$ is the same as those in Theorem \ref{thm:asymptoticnulldistribution}. 
\end{thm}

\section{Concluding remarks}
In this paper, we proposed a test for the bilateral linear hypothesis of the mean matrix in the heteroscedastic GMANOVA model where the dimensions may exceed the sample sizes. 
The proposed test is valid even for the case in which the underlying population distribution is not normal. 
In Theorem \ref{thm:rate-consistency}, we showed the rate consistency of $\hat{a}_{i,2}$, which is defined in Lemma \ref{lem:unbiased estimator}. 
This consistency was proved under the asymptotic framework that $p/N_i$ converges to a non-negative constant as $\min\{N_1,\ldots,N_g,p\}$ tends toward infinity; thus, a future study will show the consistency without assuming the convergence of $p/N_i$, to guarantee the performance for a case in which $p$ is arbitrarily larger than $N_i$.  

\section*{Acknowledgment}
The first author's research is partially supported by the Japan Society for the Promotion of Science, a Grant-in-Aid for Scientific Research (C), JSPS KAKENHI Grant Number JP18K03419, 2018-2021. 
The second author's research is partially supported by the Japan Society for the Promotion of Science, a Grant-in-Aid for Scientific Research, Young Scientists (B), JSPS KAKENHI Grant Number   JP16K16018, 2016-2020. 

\appendix
\section{Derivation of asymptotic distribution}
In this section, we show the asymptotic normality of the random variable 
\[
S = \sum_{i,j}^N \omega_{ij} 
\left(
\bs{z}_i'\bar{\bs{\Sigma}}_i^{1/2} 
\boldsymbol{\mathcal{P}}'
\boldsymbol{\mathcal{P}}
\bar{\bs{\Sigma}}_j^{1/2} \bs{z}_j
+\bs{m}_i'\bar{\bs{\Sigma}}_j^{1/2} \bs{z}_j
+\bs{z}_i' \bar{\bs{\Sigma}}_i^{1/2} \bs{m}_j
\right), 
\]
where $\bar{\bs{\Sigma}}_i$ is a
positive-definite symmetric matrix 
for $i \in \llbracket N \rrbracket$. 
Note that 
\begin{align*}
T-\mathcal{Q}
= 
\tr(
\bs{\Omega}
\boldsymbol{\mathcal{E}}'
\boldsymbol{\mathcal{P}}'
\boldsymbol{\mathcal{P}}
\boldsymbol{\mathcal{E}}
)
+ 2 \tr (
\bs{\Omega}
\boldsymbol{\mathcal{E}}'
\boldsymbol{\mathcal{P}}'
\boldsymbol{\mathcal{P}}
\bs{B}
\bs{\Theta}'\bs{A}'
)=S, 
\end{align*}
for the case in which 
\begin{equation}
\bar{\bs{\Sigma}}_j = 
\bs{\Sigma}_i, 
\quad 
j \in \{\bar{N}_i+1,\ldots,\bar{N}_i+N_i\},
\quad 
i \in \llbracket g \rrbracket. 
\label{eq:cov}
\end{equation}
To simplify the notation, hereafter, we provide a proof for the case in which $\omega_{ij} \ne 0$, $i>j$. 
Using the same derivation approach, this can be proved for a general case under which there exist $i$ and $j$ such that $\omega_{ij}=0$. 
Define 
\[
\bar{\bs{\Psi}}_i=\boldsymbol{\mathcal{P}} 
\bar{\bs{\Sigma}}_i 
\boldsymbol{\mathcal{P}}',\quad
\bar{\bs{\Upsilon}}_i=
\boldsymbol{\mathcal{P}}
\bar{\bs{\Sigma}}_i^{1/2},
\quad 
i \in \llbracket N \rrbracket. 
\]
Instead of assuming A1, A2, and A3, 
we assume 
\begin{enumerate}
\item[$\bar{\mathrm{A}}$1:] 
$ \displaystyle 
\limsup_{N \to \infty} 
\rho_N < \infty$. 
\item[$\bar{\mathrm{A}}$2:] $\displaystyle 
\frac{
\max \left\{
M_i:i=1,\ldots,6
\right\}
}{
M_7^2
} \to 0 \quad (\min \{N,p\} \to \infty)$, where 
\begin{align*}
&M_1=\sum_{i,j}^N (\tr (
\bar{\bs{\Psi}}_i
\bar{\bs{\Psi}}_j )^2, \quad 
M_2=\sum_{i,j,k}^N 
\tr (
\bar{\bs{\Psi}}_i 
\bar{\bs{\Psi}}_j
)
\tr(
\bar{\bs{\Psi}}_i 
\bar{\bs{\Psi}}_k
), \\
&M_3=
\sum_{i,j,k}^N
\sqrt{
\tr (
\bar{\bs{\Psi}}_i 
\bar{\bs{\Psi}}_j
\bar{\bs{\Psi}}_i 
\bar{\bs{\Psi}}_k
)}
\tr(\bar{\bs{\Psi}}_j 
\bar{\bs{\Psi}}_k), \\
&M_4= \sum_{i,j,k,\ell}^N  \tr (
\bar{\bs{\Psi}}_i 
\bar{\bs{\Psi}}_j
\bar{\bs{\Psi}}_k 
\bar{\bs{\Psi}}_{\ell}
), 
\quad 
M_5 = 
\sum_{i,j,k}^N 
\tr (
\bar{\bs{\Psi}}_i 
\bar{\bs{\Psi}}_j
\bar{\bs{\Psi}}_i 
\bar{\bs{\Psi}}_k
),\\
&M_6 =
\sum_{i,j,k,\ell}^N 
\sqrt{
\tr (
\bar{\bs{\Psi}}_i 
\bar{\bs{\Psi}}_j
\bar{\bs{\Psi}}_i 
\bar{\bs{\Psi}}_k
)
\tr(
\bar{\bs{\Psi}}_i 
\bar{\bs{\Psi}}_j
\bar{\bs{\Psi}}_i 
\bar{\bs{\Psi}}_{\ell}
)}, \quad 
M_7=\sum_{i,j}^N 
\tr (
\bar{\bs{\Psi}}_i 
\bar{\bs{\Psi}}_j
). 
\end{align*} 
\item[$\bar{\mathrm{A}}$3:] 
$ \displaystyle 
\frac{\sum_{i=1}^N (\bs{m}_{(i)}' \bar{\bs{\Sigma}}_i \bs{m}_{(i)})^2
}{
\mathcal{M}+(1-\mathcal{M})
(\sum_{i=1}^N \bs{m}_{(i)}' \bar{\bs{\Sigma}}_i \bs{m}_{(i)})^2
}
\to 0 \quad 
(\min \{N,p\} \to \infty)$. 
\end{enumerate}
Note that 
the assertion ``$\mathrm{A1},\mathrm{A2}, \mathrm{A3} \Rightarrow \bar{\mathrm{A}}\mathrm{1},\bar{\mathrm{A}}\mathrm{2},\bar{\mathrm{A}}\mathrm{3}$'' is true 
for the case in which $(\ref{eq:cov})$ holds; 
thus, the set of $\bar{\mathrm{A}}\mathrm{1}$, $\bar{\mathrm{A}}\mathrm{2}$, and  $\bar{\mathrm{A}}\mathrm{3}$ is milder than that of $\mathrm{A1}$, $\mathrm{A2}$, and $\mathrm{A3}$.  
The following lemma is essential for the proof; however, because it is trivial, its proof has been omitted here. 
\begin{lem}
\label{lem:Appasympnull1}
Let $F$ be a $p$-dimensional distribution that satisfies $\mathrm{D2}$. 
Assume that $\bs{z}_1$ and $\bs{z}_2$ are i.i.d. as $F$. 
Then, for the non-negative definite symmetric matrices
$\bs{A}$ and $\bs{B}$ 
and the square matrix $\bs{C}$, 
there exists a numeric constant $C_1$ such that 
\begin{align*}
&E \left[
(\bs{z}_1'\bs{A}\bs{z}_1)^2 
\right] \leq C_1 \delta (\tr (\bs{A}))^2, \\
&E \left[
\bs{z}_1'\bs{A}\bs{z}_1 
\bs{z}_1'\bs{B}\bs{z}_1
\right]
\leq C_1 \delta \tr (\bs{A}) \tr (\bs{B}), \\
&E[(\bs{z}_1'\bs{C} \bs{z}_2)^4] \leq C_1^2 \delta^2 
(\tr (\bs{C}\bs{C}'))^2, 
\end{align*}
where 
$\delta = \max \{ E[z_i^4]~:~i \in \llbracket p \rrbracket \}$. 
\end{lem}
It is observed that 
\begin{align*}
E[S] = 0, \quad 
\sigma^2 = \mathrm{Var}(S)=2 \sum_{i,j}^N \tr  (
\bar{\bs{\Psi}}_i 
\bar{\bs{\Psi}}_j
) 
+4\sum_{i=1}^N \bs{m}_{(i)}' \bar{\bs{\Sigma}}_i \bs{m}_{(i)}. 
\end{align*}
Define 
\begin{align*}
\eta_1 &= 
\frac{2}{\sigma}
\bs{z}_1' \bar{\bs{\Sigma}}_1^{1/2} \bs{m}_{(1)}, \\
\eta_i &= 
\frac{2}{\sigma}
\bs{z}_i' \bar{\bs{\Sigma}}_i^{1/2} 
\left(
\sum_{j=1}^{i-1} \omega_{ij}  
\boldsymbol{\mathcal{P}}'\boldsymbol{\mathcal{P}}  
\bar{\bs{\Sigma}}_j^{1/2} \bs{z}_j
+ \bs{m}_{(i)}
\right),
\quad i \in \{2,\ldots,N\}. 
\end{align*}
It can thus be stated that 
\[
\frac{S}{\sigma} = \sum_{i=1}^N \eta_i. 
\]
Let $\mathcal{F}_i$ be a $\sigma$-algebra generated by $\{\bs{z}_1,\ldots,\bs{z}_i\}$, 
and let $\mathcal{F}_0=\{\emptyset,\Omega\}$, 
where $\emptyset$ is the empty set and $\Omega$ is the possibility space. 
It holds that 
\[
E[\eta_i|\mathcal{F}_{i-1}] 
=0, \quad i \in \llbracket N \rrbracket. 
\]
This indicates that $\{\eta_i\}$ is martingale difference sequence, 
and $\sum_{i=1}^N E[\eta_i^2]=1$. 
From Heyde and Brown \cite{Heyde1970},  
the following inequality holds: 
\[
\sup_{x \in \mathbb{R}}
\left|
P \left(
\sum_{i=1}^N \eta_i
\leq x
\right)
- \Phi(x)
\right|
\leq 
K \left[
\sum_{i=1}^N E[\eta_i^4]
+
\mathrm{Var}
\left(
\sum_{i=1}^N \eta_i^2
\right)
\right]^{1/5}, 
\]
where $K$ is a numeric constant. 
To establish the asymptotic normality of $S/\sigma$, it is sufficient to show the following conditions:  
\begin{enumerate}
\item[C1:] $\sum_{i=1}^N E[\eta_i^4] \to 0$ as $\min \{ N,p \} \to \infty$. 
\item[C2:] $\mathrm{Var}
\left(
\sum_{i=1}^N \eta_i^2
\right) \to 0$ as $\min \{ N,p \} \to \infty$. 
\end{enumerate}
In the following subsections 
(Appendix A.1. and Appendix A.2.), 
we show $\mathrm{C1}$ and $\mathrm{C2}$. 

Summarizing the aforementioned results, we obtain the following theorem: 
\begin{thm}
Under the assumptions
$\bar{\mathrm{A}}\mathrm{1}$-$\bar{\mathrm{A}}\mathrm{3}$ 
and $\mathrm{D1}$-$\mathrm{D2}$, 
$S/\sqrt{\sigma^2}$ converges to a standard normal 
distribution as $\min \{ N, p \}$ tends to infinity. 
\end{thm}

\subsection{Proof of $\mathrm{C1}$}
In this section, we show C1. 
From the Cauchy-Schwarz inequality, 
\[
E[\eta_i^4]
\leq \frac{16}{\sigma^4}
\cdot 8 
\left(
E \left[
\left( \sum_{j=1}^{i-1} \omega_{ij} \bs{z}_i' 
\bar{\bs{\Upsilon}}_i'
\bar{\bs{\Upsilon}}_j 
\bs{z}_j
\right)^4
\right]
+
E \left[
(\bs{z}_i' \bar{\bs{\Sigma}}_i^{1/2} \bs{m}_{(i)})^4
\right]
\right). 
\]
It can then be described that 
\[
\sum_{i=1}^N E[\eta_i^4]
\leq 
128(A_1+A_2), 
\]
where 
\[
A_1 = \frac{1}{\sigma^4} \sum_{i=2}^N E \left[
\left( \sum_{j=1}^{i-1} \omega_{ij} \bs{z}_i' 
\bar{\bs{\Upsilon}}_i' 
\bar{\bs{\Upsilon}}_j 
\bs{z}_j
\right)^4
\right]>0, 
\quad 
A_2 = 
\frac{1}{\sigma^4} 
\sum_{i=1}^N 
E \left[
(\bs{z}_i' \bar{\bs{\Sigma}}_i^{1/2} \bs{m}_{(i)})^4
\right]>0. 
\]

To prove that $A_1 \to 0$, we write 
\begin{align*}
A_1 &= 
\frac{1}{\sigma^4}
\sum_{i=2}^N 
\sum_{j=1}^{i-1}
\omega_{ij}^4 
E[
(\bs{z}_1' 
\bar{\bs{\Upsilon}}_i'
\bar{\bs{\Upsilon}}_j
\bs{z}_2)^4
]\\
& \quad 
+ \frac{3}{\sigma^4}
 \sum_{i=2}^N \sum_{k,\ell}^{i-1}
\omega_{ik}^2 \omega_{i \ell}^2 
E[
\bs{z}_1' 
\bar{\bs{\Upsilon}}_i' 
\bar{\bs{\Psi}}_k 
\bar{\bs{\Upsilon}}_i
\bs{z}_1 
\bs{z}_1' 
\bar{\bs{\Upsilon}}_i' 
\bar{\bs{\Psi}}_{\ell} 
\bar{\bs{\Upsilon}}_i
\bs{z}_1 
]. 
\end{align*}
There exists a numeric constant $C_3$ such that 
\[
A_1
< 
C_3 
\rho_N^2
(A_3 + A_4), 
\]
where 
\begin{align*}
A_3 
&= \frac{
\sum_{i,j}^N 
E \left[
\left(\bs{z}_1' 
\bar{\bs{\Upsilon}}_i' 
\bar{\bs{\Upsilon}}_j 
\bs{z}_2
\right)^4
\right]
}{
M_7^2
}
>0, \\
A_4 
&= \frac{
\sum_{i,j,k}^N 
E \left[
\bs{z}_1' 
\bar{\bs{\Upsilon}}_i 
\bar{\bs{\Psi}}_j 
\bar{\bs{\Upsilon}}_i 
\bs{z}_1
\bs{z}_1' 
\bar{\bs{\Upsilon}}_i 
\bar{\bs{\Psi}}_k 
\bar{\bs{\Upsilon}}_i 
\bs{z}_1
\right]
}{
M_7^2
}>0. 
\end{align*}
Note that 
the condition $\mathrm{C1}$ is established if both $A_3$ and $A_4$ converge to 0. 
From Lemma \ref{lem:Appasympnull1}, 
\[
A_3 
< 
C_1^2 \delta^2 
\frac{
M_1
}{
M_7^2
} \to 0 \quad ( \min \{ N,p \} \to \infty), 
\]
where the convergence follows from the assumption $\bar{\mathrm{A}}$2. 
Using the same method of derivation, it holds that 
\[
A_4 < 
C_1 \delta  
\frac{
M_2
}{
M_7^2
} \to 0 \quad (\min \{ N,p \} \to \infty). 
\]

Next, we prove that $A_2 \to 0$.  
It follows from Lemma \ref{lem:Appasympnull1} that 
\begin{align*}
A_2 &\leq \frac{C_1 \delta}{\sigma^4} 
\sum_{i=1}^N 
(\bs{m}_{(i)}' \bar{\bs{\Sigma}}_i \bs{m}_{(i)})^2\\
&< 
C_1 
\frac{
\sum_{i=1}^N 
(\bs{m}_{(i)}' \bar{\bs{\Sigma}}_i \bs{m}_{(i)})^2
}{
\mathcal{M} M_7^2 
+ (1-\mathcal{M})
\left(
\sum_{i=1}^N 
\bs{m}_{(i)}' \bar{\bs{\Sigma}}_i \bs{m}_{(i)}
\right)^2
} \to 0 \quad ( \min \{ N,p \} \to \infty) 
\end{align*}
under the assumption that $\bar{\mathrm{A}}$3. 
\subsection{Proof of $\mathrm{C2}$}
It can be described that 
\[
\sum_{i=1}^N (\eta_i^2-E[\eta_i^2])
=4(B_1+B_2+B_3+2B_4), 
\]
where 
\begin{align*}
B_1 &= 
\frac{1}{2\sigma^2}
\sum_{i,j}^N 
\omega_{ij}^2 
\left\{
(\bs{z}_i' 
\bar{\bs{\Upsilon}}_i' 
\bar{\bs{\Upsilon}}_j 
\bs{z}_j)^2
- \tr 
\bar{\bs{\Psi}}_i
\bar{\bs{\Psi}}_j 
\right\}, \\
B_2 &= 
\frac{1}{\sigma^2}
\sum_{i=2}^N \sum_{j,k}^{i-1} 
\omega_{ij} \omega_{i k} 
\bs{z}_i' 
\bar{\bs{\Upsilon}}_i' 
\bar{\bs{\Upsilon}}_j 
\bs{z}_j
\bs{z}_i' 
\bar{\bs{\Upsilon}}_i' 
\bar{\bs{\Upsilon}}_k 
\bs{z}_k, \\
B_3 &= 
\frac{1}{\sigma^2}
\sum_{i=1}^N 
\left\{
(\bs{z}_i' \bar{\bs{\Sigma}}_i^{1/2} \bs{m}_{(i)})^2
- \bs{m}_{(i)}' \bar{\bs{\Sigma}}_i \bs{m}_{(i)}
\right\}, \\
B_4 &= 
\frac{1}{\sigma^2}
\sum_{i=2}^N 
\bs{z}_i' \bar{\bs{\Sigma}}_i^{1/2} \bs{m}_{(i)}
\sum_{j=1}^{i-1} 
\omega_{ij} \bs{z}_i' 
\bar{\bs{\Upsilon}}_i'  
\bar{\bs{\Upsilon}}_j 
\bs{z}_j. 
\end{align*}
The following inequality then holds.  
\[
\mathrm{Var} \left( \sum_{i=1}^N \eta_i^2 \right)
\leq 16 
E[ 4(B_1^2+B_2^2+B_3^2+4B_4^2)]
= 64 (E[B_1^2]+E[B_2^2]+E[B_3^2]+4E[B_4^2]). 
\]
Here, the expectations on the right-hand side of the equality can be described as follows: 
\begin{align*}
E[B_1^2] 
&= 
\frac{1}{2\sigma^4} 
\sum_{i,j}^N 
\omega_{ij}^4 
\{
E[(\bs{z}_1' 
\bar{\bs{\Upsilon}}_i' 
\bar{\bs{\Upsilon}}_j 
\bs{z}_2)^4]
- (\tr (
\bar{\bs{\Psi}}_i
\bar{\bs{\Psi}}_j 
))^2
\}\\
& \quad 
+ 
\frac{1}{\sigma^4} 
\sum_{i,j,k}^N 
\omega_{ij}^2 \omega_{ik}^2 
\{
E[\bs{z}_1' 
\bar{\bs{\Upsilon}}_i' 
\bar{\bs{\Psi}}_j 
\bar{\bs{\Upsilon}}_i
\bs{z}_1 
\bs{z}_1' 
\bar{\bs{\Upsilon}}_i' 
\bar{\bs{\Psi}}_k 
\bar{\bs{\Upsilon}}_i 
\bs{z}_1
] 
- \tr (
\bar{\bs{\Psi}}_i \bar{\bs{\Psi}}_j
)\tr (
\bar{\bs{\Psi}}_i
\bar{\bs{\Psi}}_k
)\}, \\
E[B_2^2] &= 
\frac{2}{\sigma^4} 
\sum_{i=2}^N \sum_{j,k}^{i-1} 
\omega_{ij}^2 \omega_{i k}^2 
E[\bs{z}_1' 
\bar{\bs{\Upsilon}}_i' 
\bar{\bs{\Psi}}_j 
\bar{\bs{\Upsilon}}_i 
\bs{z}_1 
\bs{z}_1' 
\bar{\bs{\Upsilon}}_i' 
\bar{\bs{\Psi}}_k 
\bar{\bs{\Upsilon}}_i 
\bs{z}_1
]\\
& \quad 
+ \frac{4}{\sigma^4} 
\sum_{i=3}^N \sum_{j=2}^{i-1} \sum_{k,\ell}^{j-1} 
\omega_{ik} \omega_{i \ell} \omega_{jk} \omega_{j \ell} 
\tr (
\bar{\bs{\Psi}}_i  
\bar{\bs{\Psi}}_j 
\bar{\bs{\Psi}}_k 
\bar{\bs{\Psi}}_{\ell}  
), \\
E[B_3^2] &=
\frac{1}{\sigma^4} 
\sum_{i=1}^N 
\left\{
E[(\bs{m}_{(i)}'\bar{\bs{\Sigma}}_i^{1/2} \bs{z}_1)^4]
-(\bs{m}_{(i)}' \bar{\bs{\Sigma}}_i \bs{m}_{(i)})^2
\right\}, \\
E[B_4^2] &= \frac{1}{\sigma^4}
\sum_{i=2}^N \sum_{j=1}^{i-1} 
\omega_{ij}^2 
E \left[ 
\left(\bs{z}_1' \bar{\bs{\Sigma}}_i^{1/2} \bs{m}_{(i)} 
\right)^2 
\bs{z}_1' 
\bar{\boldsymbol{\Upsilon}}_i'
\bar{\bs{\Psi}}_j 
\bar{\boldsymbol{\Upsilon}}_i
\bs{z}_1 
\right]\\
& \quad 
+\frac{2}{\sigma^4} 
\sum_{i=3}^N \sum_{j=2}^{i-1} 
\sum_{k=1}^{j-1} 
\omega_{ik} \omega_{jk} 
\bs{m}_{(i)}' 
\bar{\bs{\Sigma}}_i^{1/2}
\bar{\bs{\Upsilon}}_i' 
\bar{\bs{\Psi}}_k 
\bar{\bs{\Upsilon}}_j 
\bar{\bs{\Sigma}}_j^{1/2}
\bs{m}_{(j)}. 
\end{align*}
From Lemma \ref{lem:Appasympnull1} and assumptions 
$\bar{\mathrm{A}}$1-$\bar{\mathrm{A}}$3, 
we have 
\begin{align*}
E[B_1^2] & < 
\rho_N^2   
\frac{(C_1^2\delta^2+1) M_1+2(C_1\delta +1) M_2}{2M_7^2} \to 0 \quad (\min\{ N,p \} \to \infty), \\
E[B_2^2] & < 
\rho_N^2   
\frac{
2 C_1 \delta M_3+4M_4
}{
M_7^2} \to 0 \quad (\min \{N,p\} \to \infty), \\
E[B_3^2] &\leq
\frac{ 
(C_1 \delta + 1)
\sum_{i=1}^N (\bs{m}_{(i)}' \bar{\bs{\Sigma}}_i \bs{m}_{(i)})^2
}{
\mathcal{M}
+
(1-\mathcal{M})
(\sum_{i=1}^N \bs{m}_{(i)}' \bar{\bs{\Sigma}}_i \bs{m}_{(i)})^2
} \to 0 \quad (\min \{N,p\} \to \infty), \\
E[B_4^2] &\leq 
\rho_N 
\frac{
C_1 \delta \sqrt{
(M_1+M_2)
\sum_{i=1}^N (\bs{m}_{(i)}'\bar{\bs{\Sigma}}_i \bs{m}_{(i)})^2} 
}{
\mathcal{M}
+
(1-\mathcal{M})
M_7 \sum_{i=1}^N \bs{m}_{(i)}'\bar{\bs{\Sigma}}_i \bs{m}_{(i)}
} \\
& \quad 
+
2 \rho_N
\frac{
\sqrt{M_5+M_6} \sum_{i=1}^N 
\bs{m}_{(i)}'\bar{\bs{\Sigma}}_i \bs{m}_{(i)}
}{
\mathcal{M}
+
(1-\mathcal{M})
M_7 \sum_{i=1}^N \bs{m}_{(i)}'\bar{\bs{\Sigma}}_i \bs{m}_{(i)}
}
\to 0 \quad ( \min \{N,p\} \to \infty), 
\end{align*}
which imply that $\mathrm{C2}$ is satisfied. 

\section{Rate consistency}
In this section, we prove the rate consistency stated in Theorem \ref{thm:rate-consistency}. 
It is sufficient to show that 
\begin{align*}
&\frac{\hat{a}_{1,2}}{a_{1,2}}
\overset{P}{\to} 1 \quad 
(\min\{N_1,p\} \to \infty~\mbox{and}~p/N_1 \to c \in [0,\infty)), \\ 
&\frac{\hat{b}_{12}}{b_{12}}
\overset{P}{\to} 1 \quad 
(\min\{N_1,N_2,p\} \to \infty
), 
\end{align*}
where 
\begin{align*}
\hat{a}_{1,2}
= \nu_1 \tr (\bs{S}_1^2) - \nu_2 ( \tr (\bs{S}_1))^2 
- \nu_3 Q_1, \quad 
\hat{b}_{12}
= \tr (\bs{S}_1 \bs{S}_2). 
\end{align*}
Here, 
\begin{align*}
\nu_1 = \frac{(N_1-k_1)^2 \tau_{1,2} - \tau_{1,1}^2}{(N_1-k_1) \tau_{1,3}}, \quad 
\nu_2 = \frac{(N_1-k_1) \tau_{1,2} - \tau_{1,1}^2}{(N_1-k_1) \tau_{1,3}}, \quad 
\nu_3 = \frac{(N_1-k_1-1) \tau_{1,1}}{(N_1-k_1) \tau_{1,3}}. 
\end{align*}
Let 
\[
\bs{\Upsilon}_i =  \boldsymbol{\mathcal{P}}^{1/2} 
\bs{\Sigma}_i, \quad i=1,2. 
\]

\subsection{Proof of the rate consistency for $a_{1,2}$}
To show the rate consistency, we use the following lemmas (Lemma \ref{lem:consistency2}-\ref{lem:consistency4}): 
\begin{lem}
\label{lem:consistency2} 
The following four results hold. 
\[
\lim_{N_1 \to \infty} \nu_1 = 1, \quad 
\nu_2 = O(N_1^{-2}), \quad 
\nu_3 = O(N_1^{-1}), 
\quad 
\nu_1-(N_1-k_1) \nu_3 = O(N_1^{-1}).  
\]
\end{lem}

\begin{lem}
\label{lem:consistency3}
Under the assumptions 
$\mathrm{D1}$ and $\mathrm{D2}$, 
the following probability convergences hold: 
\begin{enumerate}
\item[$\mathrm{(i)}$] $
\displaystyle 
\frac{\tr (\bs{S}_1)}{\tr( 
\bs{\Psi}_1 
)
} \overset{P}{\to} 1$ 
as $\min\{N_1,p\}$ tends toward infinity. 
\item[$\mathrm{(ii)}$] $
\displaystyle 
\frac{
\tr (\bs{S}_1^2)
}{
\tr 
(
\bs{\Psi}_1^2)
}-\frac{1}{(N_1-k_1)^2} 
\frac{ \tr (\bs{Z} 
\bs{\Upsilon}_1'\bs{\Upsilon}_1
\bs{Z}')^2
}{ 
\tr 
(
\bs{\Psi}_1^2)}
\overset{P}{\to} 0
$ 
under the asymptotic framework that 
$p/N_1$ converges to a non-negative constant 
as $\min\{N_1,p\}$ tends toward infinity. 
\item[$\mathrm{(iii)}$] $
\displaystyle 
\frac{Q_1}{ p \tr 
(
\bs{\Psi}_1^2)}
-\frac{1}{N_1-k_1} 
\frac{ \tr
((\bs{Z} 
\bs{\Upsilon}_1'\bs{\Upsilon}_1
\bs{Z}')
\odot
(\bs{Z} 
\bs{\Upsilon}_1' \bs{\Upsilon}_1
\bs{Z}'))
}{ p \tr 
(\bs{\Psi}_1^2) }
\overset{P}{\to} 0
$ 
under the asymptotic framework that 
$p/N_1$ converges to a non-negative constant 
as $\min\{N_1,p\}$ tends toward infinity. 
\end{enumerate}
Here, $\bs{Z}=(\bs{z}_1,\ldots,\bs{z}_{N_1})'
=(\bs{z}_1^{(1)},\ldots,\bs{z}_{N_1}^{(1)})
$. 
\end{lem}

\begin{lem}
\label{lem:consistency4}
Under the assumptions $\mathrm{D1}$ and $\mathrm{D2}$, 
the following probability of convergences holds: 
\begin{enumerate}
\item[$\mathrm{(i)}$] $
\displaystyle 
\frac{1}{N_1^2} 
\sum_{i=1}^{N_1}
\frac{(\bs{z}_i' 
\bs{\Upsilon}_1' \bs{\Upsilon}_1 
\bs{z}_i)^2
}{ ( \tr 
(
\bs{\Psi}_1 
) 
)^2 }
\overset{P}{\to} 0
$ 
as $\min\{N_1,p\}$ tends toward infinity. 
\item[$\mathrm{(ii)}$] $
\displaystyle 
\frac{1}{N_1(N_1-1)} 
\sum_{i,j}^{N_1} 
\frac{
(\bs{z}_i' 
\bs{\Upsilon}_1' \bs{\Upsilon}_1 
\bs{z}_j)^2
}{ 
\tr (
\bs{\Psi}_1^2)
}
\overset{P}{\to} 1
$ 
as $\min\{N_1,p\}$ tends toward infinity. 
\end{enumerate}
Here, $(\bs{z}_1,\ldots,\bs{z}_{N_1})=(\bs{z}_1^{(1)},\ldots,\bs{z}_{N_1}^{(1)})$. 
\end{lem}
\noindent
Proofs of these lemmas are simple but tedious; therefore, they have been omitted (they are provided in the supplementary material). 

It can be described that 
\begin{align*}
\frac{
\hat{a}_{1,2}
}{
\tr 
(
\bs{\Psi}_1^2)
}
&= \nu_1 \frac{\tr (\bs{S}_1^2)}{
\tr 
(
\bs{\Psi}_1^2
)
} 
- \frac{p}{N_1^2} (N_1^2 \nu_2) 
\frac{
(\tr 
(\bs{\Psi}_1)
)^2}{ p \tr (
\bs{\Psi}_1^2)
}
\left(
\frac{ \tr (\bs{S}_1) }{ 
\tr (
\bs{\Psi}_1
)
}
\right)^2 
- N_1 \nu_3 \frac{p}{N_1} \frac{Q_1}{p 
\tr (
\bs{\Psi}_1^2)
}. 
\end{align*}
Applying Lemma \ref{lem:consistency2}-\ref{lem:consistency3}, 
\begin{align*}
&\frac{\hat{a}_{1,2}}{
\tr 
(
\bs{\Psi}_1^2
)
}
-\left\{
\frac{\nu_1}{(N_1-k_1)^2}
\frac{\tr ((\bs{Z} 
\bs{\Upsilon}_1' \bs{\Upsilon}_1 
\bs{Z}' )^2)}{\tr 
(\bs{\Psi}_1^2)
}
-\frac{\nu_3}{N_1-k_1} 
\frac{ \tr ((\bs{Z}
\bs{\Upsilon}_1' \bs{\Upsilon}_1 
\bs{Z}') \odot 
(\bs{Z} 
\bs{\Upsilon}_1'\bs{\Upsilon}_1 
\bs{Z}')) }
{ 
\tr (
\bs{\Psi}_1^2)
}
\right\}
\overset{P}{\to} 0 
\end{align*}
under the asymptotic framework that $p/N_1$ converges to a non-negative constant as $\min\{N_1, p\} \to \infty$. 
Here, the expression in the braces can be expressed as follows. 
\begin{equation}
\frac{\nu_1}{(N_1-k_1)^2}
\sum_{i,j}^{N_1} 
\frac{(\bs{z}_i' 
\bs{\Upsilon}_1' \bs{\Upsilon}_1 
\bs{z}_j)^2
}{
\tr (
\bs{\Psi}_1^2
)
}
+ \left\{
\frac{\nu_1}{(N_1-k_1)^2}
-\frac{\nu_3}{N_1-k_1}
\right\}
\sum_{i=1}^{N_1} 
\frac{(\bs{z}_i' 
\bs{\Upsilon}_1' \bs{\Upsilon}_1 
\bs{z}_i)^2
}{
\tr 
(
\bs{\Psi}_1^2
)
}, \label{eq:unbias2}
\end{equation} 
which converges to 1 in probability, 
where the convergence is followed from Lemma \ref{lem:consistency2}, Lemma \ref{lem:consistency4}, and 
$0 \leq 
( \tr (
\bs{\Psi}_1
))^2
/\{
p 
\tr (
\bs{\Psi}_1^2
 )
\}
\leq 1
$. 
Summarizing the aforementioned results, we can see that $\hat{a}_{1,2}/a_{1,2}$ converges to 1 in probability.

\subsection{Proof of rate consistency for $b_{12}$}
Let us put $\bs{Z}_1=(\bs{z}_1,\ldots,\bs{z}_{N_1})'
=(\bs{z}_1^{(1)},\ldots,\bs{z}_{N_1}^{(1)})
$ and 
$\bs{Z}_2=(\bs{z}_{N_1+1},\ldots,\bs{z}_{N_1+N_2})'
=(\bs{z}_1^{(2)},\ldots,\bs{z}_{N_2}^{21)})
$. 
To show the rate consistency, we use the following lemmas. 

\begin{lem}
\label{lem:consistency7}
Under the assumptions $\mathrm{D1}$ and $\mathrm{D2}$, 
\[
\frac{
\tr (\bs{S}_1 \bs{S}_2)
}{
\tr 
(
\bs{\Psi}_1
\bs{\Psi}_2
)
}
-\frac{1}{(N_1-k_1)(N_2-k_2)} 
\frac{ \tr (\bs{Z}_1
\bs{\Upsilon}_1' \bs{\Upsilon}_2 
\bs{Z}_2'
\bs{Z}_2 
\bs{\Upsilon}_2' \bs{\Upsilon}_1
\bs{Z}_1')
}{ 
\tr 
(
\bs{\Psi}_1
\bs{\Psi}_2
)
}
\overset{P}{\to} 0
\]
as $\min\{N_1N_2,p\}$ tends toward infinity. 
\end{lem}
\begin{lem}
\label{lem:consistency8}
Under the assumptions $\mathrm{D1}$ and $\mathrm{D2}$, 
\[
\frac{
\sum_{i=1}^{N_1} \sum_{j=1}^{N_2}
(\bs{z}_i' 
\bs{\Upsilon}_1' \bs{\Upsilon}_2 
\bs{z}_j
)^2
}{
N_1 N_2 
\tr 
(
\bs{\Psi}_1\bs{\Psi}_2
)
} 
\overset{P}{\to} 1 
\]
as $\min\{N_1N_2,p\}$ tends toward infinity. 
\end{lem}
\noindent
Lemma \ref{lem:consistency7} can be proved using the derivation method applied
to prove Lemma \ref{lem:consistency3} (ii) and 
Lemma \ref{lem:consistency8} can be proved using the derivation method used 
to prove Lemma \ref{lem:consistency4} (ii); 
these proofs have, therefore, been omitted from this paper. 

From Lemma \ref{lem:consistency7}-\ref{lem:consistency8}, 
\begin{align*}
\frac{
\tr (\bs{S}_1 \bs{S}_2)
}{
\tr (
\bs{\Psi}_1
\bs{\Psi}_2
)
}-1
&=
\frac{
\tr (\bs{S}_1 \bs{S}_2)
}{
\tr 
(
\bs{\Psi}_1
\bs{\Psi}_2
)
}
-\frac{1}{(N_1-k_1)(N_2-k_2)} 
\frac{ 
\tr (\bs{Z}_1 
\bs{\Upsilon}_1'\bs{\Upsilon}_2 
\bs{Z}_2'
\bs{Z}_2 
\bs{\Upsilon}_2' \bs{\Upsilon}_1
\bs{Z}_1')
}{ 
\tr (
\bs{\Psi}_1
\bs{\Psi}_2
)
}\\
& \quad 
+
\frac{k_2N_1+k_1N_2-k_1k_2}{(N_1-k_1)(N_2-k_2)}
\frac{
\sum_{i=1}^{N_1} \sum_{j=1}^{N_2}
(\bs{z}_i' 
\bs{\Upsilon}_1' \bs{\Upsilon}_2
\bs{z}_j
)^2
}{
N_1 N_2 
\tr 
(
\bs{\Psi}_1
\bs{\Psi}_2
)
}\\
& \quad 
+
\frac{
\sum_{i=1}^{N_1} \sum_{j=1}^{N_2}
(\bs{z}_i' 
\bs{\Upsilon}_1' \bs{\Upsilon}_2
\bs{z}_j
)^2
}{
N_1 N_2 
\tr 
(
\bs{\Psi}_1
\bs{\Psi}_2
)
} -1 \\
& \overset{P}{\to} 0. 
\end{align*}

\section*{Supplementary materials}
In the supplementary material, we prove Lemma \ref{lem:unbiased estimator} 
and Lemma \ref{lem:consistency2}-\ref{lem:consistency4}. 
In addition, specific results contained in the proposed class of tests are reviewed, and the results of numerical evaluations conducted through various finite-sample simulation scenarios, along with an example using real data are also reported.  

\bibliographystyle{plain}
\bibliography{YHAP}

\begin{thebibliography}{10}

\bibitem{Anderson2003}
T.~W. Anderson.
\newblock {\em An Introduction to Multivariate Statistical Analysis}.
\newblock Wiley Series in Probability and Statistics. Wiley-Interscience [John
  Wiley \& Sons], Hoboken, NJ, third edition, 2003.

\bibitem{Bai2018}
Zhidong Bai, Kwok~Pui Choi, and Yasunori Fujikoshi.
\newblock Limiting behavior of eigenvalues in high-dimensional {MANOVA} via
  {RMT}.
\newblock {\em Ann. Statist.}, 46(6A):2985--3013, 2018.

\bibitem{Bai1996}
Zhidong Bai and Hewa Saranadasa.
\newblock Effect of high dimension: by an example of a two sample problem.
\newblock {\em Statist. Sinica}, 6(2):311--329, 1996.

\bibitem{Cai2014a}
T.~Tony Cai, Weidong Liu, and Yin Xia.
\newblock Two-sample test of high dimensional means under dependence.
\newblock {\em J. R. Stat. Soc. Ser. B. Stat. Methodol.}, 76(2):349--372, 2014.

\bibitem{Cai2014b}
T.~Tony Cai and Yin Xia.
\newblock High-dimensional sparse {MANOVA}.
\newblock {\em J. Multivariate Anal.}, 131:174--196, 2014.

\bibitem{Chen2019}
Song~Xi Chen, Jun Li, and Ping-Shou Zhong.
\newblock Two-sample and {ANOVA} tests for high dimensional means.
\newblock {\em Ann. Statist.}, 47(3):1443--1474, 2019.

\bibitem{Chen2010a}
Song~Xi Chen and Ying-Li Qin.
\newblock A two-sample test for high-dimensional data with applications to
  gene-set testing.
\newblock {\em Ann. Statist.}, 38(2):808--835, 2010.

\bibitem{Fujikoshi2010}
Yasunori Fujikoshi, Vladimir~V. Ulyanov, and Ryoichi Shimizu.
\newblock {\em Multivariate Statistics: High-Dimensional and Large-Sample
  Approximations}.
\newblock Wiley Series in Probability and Statistics. John Wiley \& Sons, Inc.,
  Hoboken, NJ, 2010.

\bibitem{Ghosh2016}
Anil~K. Ghosh and Munmun Biswas.
\newblock Distribution-free high-dimensional two-sample tests based on
  discriminating hyperplanes.
\newblock {\em TEST}, 25(3):525--547, 2016.

\bibitem{Heyde1970}
C.~C. Heyde and B.~M. Brown.
\newblock On the departure from normality of a certain class of martingales.
\newblock {\em Ann. Math. Statist.}, 41:2161--2165, 1970.

\bibitem{Himeno2014}
Tetsuto Himeno and Takayuki Yamada.
\newblock Estimations for some functions of covariance matrix in high dimension
  under non-normality and its applications.
\newblock {\em J. Multivariate Anal.}, 130:27--44, 2014.

\bibitem{Jana2017}
Sayantee Jana, Narayanaswamy Balakrishnan, Dietrich von Rosen, and Jemila~Seid
  Hamid.
\newblock High dimensional extension of the growth curve model and its
  application in genetics.
\newblock {\em Stat. Methods Appl.}, 26(2):273--292, 2017.

\bibitem{Muirhead1982}
Robb~J. Muirhead.
\newblock {\em Aspects of Multivariate Statistical Theory}.
\newblock John Wiley \& Sons, Inc., New York, 1982.
\newblock Wiley Series in Probability and Mathematical Statistics.

\bibitem{Srivastava2002}
Muni~S. Srivastava.
\newblock {\em {Methods of Multivariate Statistics}}.
\newblock Wiley Series in Probability and Statistics. Wiley-Interscience [John
  Wiley \& Sons], New York, 2002.

\bibitem{Srivastava2013}
Muni~S. Srivastava and Tatsuya Kubokawa.
\newblock {Tests for multivariate analysis of variance in high dimension under
  non-normality}.
\newblock {\em Journal of Multivariate Analysis}, 115:204--216, 2013.

\bibitem{Srivastava2017}
Muni~S. Srivastava and Martin Singull.
\newblock Test for the mean matrix in a growth curve model for high dimensions.
\newblock {\em Comm. Statist. Theory Methods}, 46(13):6668--6683, 2017.

\bibitem{Rosen2018}
Dietrich von Rosen.
\newblock {\em Bilinear Regression Analysis: An Introduction}, volume 220 of
  {\em Lecture Notes in Statistics}.
\newblock Springer, Cham, 2018.

\bibitem{Wang2015}
Lan Wang, Bo~Peng, and Runze Li.
\newblock A high-dimensional nonparametric multivariate test for mean vector.
\newblock {\em J. Amer. Statist. Assoc.}, 110(512):1658--1669, 2015.

\bibitem{Wang2019}
Wei Wang, Nan Lin, and Xiang Tang.
\newblock Robust two-sample test of high-dimensional mean vectors under
  dependence.
\newblock {\em Journal of Multivariate Analysis}, 169:312--329, 2019.

\bibitem{Yamada2015}
Takayuki Yamada and Tetsuto Himeno.
\newblock Testing homogeneity of mean vectors under heteroscedasticity in
  high-dimension.
\newblock {\em J. Multivariate Anal.}, 139:7--27, 2015.

\bibitem{Yamada2012}
Takayuki Yamada and Tetsuro Sakurai.
\newblock Asymptotic power comparison of three tests in gmanova when the number
  of observed points is large.
\newblock {\em Statistics \& Probability Letters}, 82(3):692--698, 2012.

\bibitem{Zhou2017}
Bu~Zhou, Jia Guo, and Jin-Ting Zhang.
\newblock High-dimensional general linear hypothesis testing under
  heteroscedasticity.
\newblock {\em J. Statist. Plann. Inference}, 188:36--54, 2017.

\bibitem{Zhou2020}
Bu~Zhou, Jia Guo, and Jin-Ting Zhang.
\newblock An $\boldsymbol{L^{2}}$-norm based test for high-dimensional two-way
  manova.
\newblock {\em Scientia Sinica Mathematica}, 50(5):729--750, 2020.

\end{thebibliography}

\end{document}